\documentstyle[multicol,aps,psfig]{revtex}  
\begin{document}  
\newcommand{\vk}{{\vec k}}  
\newcommand{\vK}{{\vec K}}   
\newcommand{\vb}{{\vec b}}   
\newcommand{{\vp}}{{\vecp}}   
\newcommand{{\vq}}{{\vec q}}   
\newcommand{\vQ}{{\vec Q}}  
\newcommand{\vx}{{\vec x}}  
\newcommand{\vh}{{\hat{v}}}  
\newcommand{\tr}{{{\rm Tr}}}   
\newcommand{\beq}{\begin{equation}}  
\newcommand{\eeq}[1]{\label{#1} \end{equation}}   
\newcommand{\half}{{\textstyle\frac{1}{2}}}   
\newcommand{\lton}{\mathrel{\lower.9ex \hbox{$\stackrel{\displaystyle  
<}{\sim}$}}}   
\newcommand{\ee}{\end{equation}}  
\newcommand{\ben}{\begin{enumerate}}   
\newcommand{\een}{\end{enumerate}}  
\newcommand{\bit}{\begin{itemize}}   
\newcommand{\eit}{\end{itemize}}  
\newcommand{\bc}{\begin{center}}   
\newcommand{\ec}{\end{center}}  
\newcommand{\bea}{\begin{eqnarray}}   
\newcommand{\eea}{\end{eqnarray}}  
\newcommand{\beqar}{\begin{eqnarray}}   
\newcommand{\eeqar}[1]{\label{#1}  
\end{eqnarray}}   
\newcommand{\gton}{\mathrel{\lower.5ex \hbox{$\stackrel{> } 
 {\scriptstyle \sim}$}}} 
 
\title{Transverse Expansion and  High $p_T$ Azimuthal Asymmetry at RHIC}   
  
\author{Miklos~Gyulassy$^{1}$, Ivan~Vitev$^{1}$, 
 Xin-Nian Wang$^2$, and Pasi Huovinen$^2$} 
  
\address{$^1$  Department of Physics, Columbia University,   
         538 W. 120-th Street, New York, NY 10027\\  
         $^2$ Nuclear Science Division, Lawrence Berkeley National Lab,  
       Berkeley, CA 94720}  
  
\maketitle  
  
\begin{abstract}  
Rapid 3+1D transverse plus Bjorken collective expansion in $A+A$ 
collisions at ultra-relativistic energies is shown to reduce 
substantially the azimuthal asymmetry resulting from jet quenching. 
 While the azimuthal asymmetry in non-central collisions, 
$v_2(p_T>2\;\rm{GeV/c})\sim 0.15$ reported  
by  STAR at RHIC, can be accounted for by spatially anisotropic jet 
energy loss through a 1+1D expanding gluon plasma 
with $dN^g/dy\sim 1000$, we show that if rapid transverse collective 
expansion of the plasma is assumed, then the asymmetry 
due to jet quenching may 
be reduced  below the observed level. 
Possible implications of this effect are discussed.  
\end{abstract}  
 
\begin{multicols}{2}   
    
\section{Introduction}   
Preliminary STAR data from RHIC~\cite{Snellings,starv2,Adler:2001nb,reanv2} 
suggest a large transverse azimuthal asymmetry $v_2\sim 0.10-0.15$ at   
moderate high $2 < p_T < 5$~GeV/c.  
In Refs.~\cite{gvw,wangv2} some of us proposed that the saturation of  
$v_2(p_T)$ could  be interpreted as a 
manifestation of asymmetric jet energy loss that arises  
due to the different jet propagation 
path lengths in non-central (${\bf b} > 0\; {\rm fm}$) collisions 
of heavy ions. We applied the GLV~\cite{glvprl,glv2,glv1b,glv1a} energy 
loss formalism 
to compute the asymmetry and found that under the approximation that 
the initial spatial anisotropy of the plasma 
(generated by intersecting sharp cylinders)
did not change with time 
the observed high $p_T$ $v_2$ could be explained by asymmetric jet quenching. 
The  initial gluon rapidity density when extrapolated to central 
collisions was estimated to be $dN_g/dy({\bf b}=0\; {\rm fm}) \sim 1000$. 
The implied large gluon density, 
$\rho_g(\tau,{\bf b})\approx dN^g/dy({\bf b})/(\tau A_\perp({\bf b}))$, 
was also found to be consistent with the azimuthally averaged jet quenching 
pattern of $\pi^0$ at $p_T > 2$~GeV/c reported by 
PHENIX~\cite{pi0,bill,phenqnch}. 
 
In Ref.~\cite{gvw} the non-perturbative low $p_T$  component  of 
the hadronic spectra was parametrized in terms of hydrodynamic 
solutions~\cite{pasi,Hirano:2001yi} that fit well the 
observed $p_T < 2 $ GeV/c flavor dependent inclusive distributions  
for pions and nucleons~\cite{Adler:2001nb,Adler:2001yq}. 
We ignored in Ref.~\cite{gvw}, however, the additional rapid dilution 
and time dependent spatial asymmetry of the plasma that is predicted by 
hydrodynamic transverse flow. 
Boosted thermal model 
fits~\cite{Xu:2001zj,Broniowski:2001we,gv,gvproc,gvPetproc}  
to the low $p_T$ data suggest that 
the collective transverse flow velocities  
may be quite high at RHIC with $v_T\sim 0.6$. 
The purpose of this letter is to compute the effect of such 3+1D  
expansion on the azimuthal asymmetry of the jet quenching pattern. 
We find that while in extreme scenarios, assuming instantaneous 
transverse expansion
of the medium, the geometric anisotropy is strongly reduced, the
overall magnitude of the mean energy loss $\Delta E$ is much less 
sensitive for flow velocities up to $v_T=0.8$. This can be verified 
for a variety of transverse density profiles as well as a full hydrodynamic
calculation~\cite{pasi,pasiold,pasinew}.  We also speculate about 
different non-perturbative 
effects on moderate $p_T$  
meson and baryon production that could lead to different 
flavor dependent high $p_T$ anisotropy behavior of hadron yields.

\section{Energy loss of jets in transversely expanding medium} 
 
The  explicit solution to the problem of energy loss of a  
hard ultra-relativistic jet ($c=1, z=\tau$)  
produced inside a hot and dense medium at position $\vec{x}_0$ and propagating 
in direction $\hat{v}$ was obtained in~\cite{glv2}  
to all orders in opacity  
\begin{equation}  
\chi =\int_{\tau_0}^\infty d\tau \;   
\sigma(\tau) \rho(\vx_0 + \vh (\tau-\tau_0),\tau)\; \; . 
\label{opac}  
\end{equation}  
It was shown that for gluon plasma of transverse thickness  
on the order of the nuclear size and density $\rho \leq 50$~fm$^{-3}$  
(relevant at the currently available heavy ion collider energies)  
the dominant contribution  comes from the first order in the opacity  
expansion~\cite{glvprl,glv2,zakh}. 
In the case of 1+1D Bjorken longitudinal expansion with initial plasma 
density $\rho_0=\rho(z_0\equiv\tau_0)$ and formation time  
$z_0\equiv\tau_0$, i.e.   
\beq 
\rho(z=\tau)=\rho_0\left(\frac{\tau_0}{\tau}\right)^\alpha \;\;, 
\eeq{bjork} 
it is possible to obtain  a closed form analytic formula~\cite{gvw}    
(under the strong asymptotic no kinematic bounds assumption) 
for the energy loss due to 
the dominant first order term~\cite{glvprl,glv2}. For a hard jet  
penetrating the quark-gluon plasma 
\beqar 
\frac{d \Delta E^{(1)}}{dx} &=&  
   \frac{2C_R\alpha_s}{\pi} E   
\int^\infty_{\tau_0} \frac{d \tau}{\lambda(\tau)}\; f(Z(x,\tau)), 
\eeqar{de11}  
where $x\simeq \omega/E$ is the momentum fraction of the 
radiated gluon and  
the formation physics function $f(Z(x,\tau))$ is defined in~\cite{gvw} 
to be 
\beqar 
f(x,\tau)&=& \int_0^\infty \frac{du}{u(1+u)}   
\left[ 1 -   \cos \left (\,u Z(x,\tau) \right) \right]  \;. 
\eeqar{fz} 
With $Z(x,\tau)=(\tau-\tau_0)\mu^2(\tau)/2 x E$ being the local formation physics  
parameter,  two simple analytic limits of 
Eq.~(\ref{fz}) can be obtained. For $x \gg x_c =  
\mu(\tau_0)^2 \tau_0^{ \frac{2\alpha}{3}}  
L^{1- \frac{2\alpha}{3} }/2E 
= L\mu^2(L)/2 E$, in which case  
the formation length is large compared to the size of the medium,  
the small $Z(x,\tau)$ limit applies leading to 
$f(Z) \approx \pi Z / 2 $. The interference pattern along the 
gluon path becomes important and accounts for the the 
non-trivial dependence of the energy loss on $L$.   
When $x \ll x_c$, i.e. the formation length  
is small compared to the plasma thickness,  
one gets $f(Z) \approx \log Z$.  
In the $x \gg x_c$ limit~\cite{gvw} 
the radiative spectrum  
Eq.~(\ref{de11}) becomes 
\beqar 
\frac{d \Delta E^{(1)}_{x\gg x_c}}{dx} \approx  
\frac{C_R \alpha_s}{2(2-\alpha)}  
\frac{\mu(\tau_0)^2\tau_0^\alpha L^{2-\alpha}}{\lambda(\tau_0)}  
\frac{1}{x} \;.  
\eeqar{dIdxxbig}  
The mean  energy loss (to first order in $\chi$) integrates to 
\beqar 
\Delta E^{(1)} &=&  \frac{C_R \alpha_s}{2(2-\alpha)}  
\frac{\mu(\tau_0)^2\tau_0^\alpha L^{2-\alpha} }{\lambda(\tau_0)}  \times  
\nonumber  
\\[1ex] 
&&   \times \; 
\left( 
 \log \frac{2 E}{\mu(\tau_0)^2 \tau_0^{ \frac{2\alpha}{3}}  
L^{1- \frac{2\alpha}{3} }}   
+  \cdots \right)  \;\;.   
\eeqar{totde} 
The logarithmic enhancement with energy comes from the  
$x_c<x<1$ region~\cite{glvprl,glv2}. 
In the case of sufficiently  large jet energies  
($E\rightarrow \infty$) this term dominates. For parton energies   
$<20$~GeV, however, corrections to this leading $\log 1/x_c$  expression 
that can be exactly evaluated numerically from the GLV expression 
and are found to be 
comparable in size. The  effective $\Delta E/E$ in this energy range  is 
approximately constant~\cite{levai}. It was also recently shown that 
the effect of gluon absorption is negligible for jet energies 
$E \geq 5 \mu \sim 3$~GeV~\cite{WW}. 
 
In the above expression, transverse expansion can only be very crudely modeled 
taking $\alpha > 1$. To derive an improved  analytic expression 
taking transverse flow into account, 
we consider next an asymmetric expanding sharp {\em elliptic}  
density profile the surface of which is defined by  
\beq  
\frac{x^2}{(R_x+v_x \tau)^2} +\frac{y^2}{(R_y+v_y \tau)^2} = 1 \;\;. 
\eeq{ellip}  
The area of this elliptic transverse profile increases with time , $\tau$,  
as  
\beq  
A_\perp(\tau)=\pi(R_x+v_x \tau) (R_y+v_y \tau) \;\;. 
\eeq{area}  
Consider the plasma density seen by a jet  
in direction  
$\phi_0$ starting from ${\vec{\bf x}_0}=(x_0,y_0)$ inside the ellipse  
with a specified initial density $\tau_0\rho_0= 1 / (\pi R_xR_y)\, dN^g/dy$.  
The density along its path is  
\beqar  
\rho(\tau,\phi_0;x_0,y_0)&=&  
\frac{1}{\pi}  
\frac{dN^g}{dy}\left(\frac{1}{\tau}\right) \left(\frac{1}{R_x+v_x \tau}\right)  
 \left(\frac{1}{R_y+v_y \tau}\right)  
 \nonumber \\[1ex]  
&\;& \hspace{-1.5cm}\theta\left(1-\frac{(x_0+\tau \cos \phi_0)^2}{(R_x+v_x \tau)^2}   
+\frac{(y_0+\tau \sin \phi_0)^2}{(R_y+v_y \tau)^2}\right) \;.  
\eeqar{rhoex}

Analytic expressions can be obtained  only for asymptotic jet energies  
when the kinematic boundaries can be ignored. 
In that case Eq.~(\ref{de11}) reduces to  
\beqar  
\frac{d \Delta E^{(1)}}{dx} 
&\approx& \frac{C_R\alpha_s}{2x} \int^{\infty}_{0}   
d \tau \,\frac{\mu^2(\tau)}{\lambda_g(\tau)}  
\; \theta \left(\frac{2xE}{\tau\mu^2(\tau)}-1\right)
\;\;,  
\eeqar{ndif11}   
where we kept only  the dominant $\log 1/x_c$, $Z < 1$ regime.  
By integrating over $x$ the total energy loss is  
for this density profile azimuthally dependent: 
\beqar  
\Delta E^{(1)}(\phi_0)   
  &\approx& \frac{9\pi C_R\alpha_s^3}{4}\int^\infty_{0} d \tau \,  
\rho(\tau,\phi_0)\;\tau \, \log \frac{2E}{\tau\mu(\tau)^2}\;\; ,  
\eeqar{de11int}   
which is a linear (mod log) weighed line integral over the   
asymmetric density. This integral is still  convergent in spite of  
the singularity of the density at $\tau=0$.

We consider in more detail the analytically tractable case of a sharp 
expanding elliptic cylinder.
We approximate the assumed $\phi_0$ independent screening  
$\mu(\tau)\approx g T(\tau) = 2 (\rho(\tau)/2)^{1/3}$ since  
$g \simeq 2$ and $\rho = (16 \zeta(3)/ \pi^2) T^3 \simeq 2 T^3 $ for  
gluon plasma.   
We define   
$\tau(\phi_0)$ as the  escape time to reach the expanding elliptic surface from  
an initial point $\vec{\bf x}_0$ in the azimuthal direction $\phi_0$:  
 \beq  
\frac{(x_0+\tau \cos \phi_0)^2}{(R_x+v_x \tau)^2}   
+\frac{(y_0+\tau \sin \phi_0)^2}{(R_y+v_y \tau)^2}=1  
\;\; . \eeq{escape}  
We take $\omega(\phi_0)=2\, \tau(\phi_0) \,( \rho(\tau(\phi_0))/2 )^{2/3}$  
to estimate an upper bound on the logarithmic enhancement factor.  
Performing the remaining integrals one gets: 
\beqar  
\Delta E^{(1)}(\phi_0)   
  &\approx& \frac{9 C_R\alpha_s^3}{4}\frac{dN^g}{dy}  
\; \log \frac{E}{\omega(\phi_0)} \, \times 
\nonumber \\  
&\;& \hspace{-.5cm} \times \, \int^\infty_{0} d \tau \,  
\left(\frac{1}{R_x+v_x \tau}\right)  
 \left(\frac{1}{R_y+v_y \tau }\right)\nonumber \\[1ex]  
&\;& \hspace{-.5cm}
\theta\left( 1 - \frac{(x_0+\tau \cos \phi_0)^2}{(R_x+v_x \tau)^2}   
+ \frac{(y_0+\tau \sin \phi_0)^2}{(R_y+v_y \tau)^2}\right)
\nonumber\\[1ex]  
&=& \frac{9}{4} \, \frac{ C_R\alpha_s^3}{R_x R_y}\frac{dN^g}{dy}  
\; \frac{\log  
\frac{1 + a_x \tau(\phi_0)}{1+ a_y \tau(\phi_0)}}{a_x-a_y}  
\; \log \frac{E}{\omega(\phi_0)} 
\;\; , 
\eeqar{deaz}   
where $a_x=v_x/R_x, a_y=v_y/R_y$. 
This expression is the central result of this paper 
and provides a simple analytic  generalization 
that interpolates between pure Bjorken 1+1D expansion for small 
$a_{x,y} \tau$, and 3+1D expansion at large $a_{x,y} \tau$. 
 
In the special case of pure Bjorken (longitudinal) expansion 
with $v_x=v_y=0$  
\beqar  
\Delta E^{(1)}_{Bj}(\phi_0)&=&   
\frac{9C_R\alpha_s^3}{4 R_x R_y}   
\frac{dN^g}{dy}  \, \tau(\phi) \,\log \frac{E}{\omega(\phi_0)}  
\; . 
\eeqar{debj} 
In this case, the energy loss depends  {\em linearly} on  $\tau(\phi)$. 
 
Another special case is azimuthally {\em isotropic} expansion 
with $R_x=R_y=R$ and $v_x=v_y=v_T$. Taking also 
the longitudinal Bjorken expansion into account leads in this case to 
\beqar  
\Delta E^{(1)}_{3D}(\phi_0)&=&   
\frac{9}{4} \frac{C_R\alpha_s^3}{R^2}   
\frac{dN^g}{dy}  \, \frac{\tau(\phi_0)}{1+v_T\tau(\phi_0)/R}  
\,\log \frac{E}{\omega(\phi_0)} 
\; .   
\eeqar{debj1}  
We note that for a jet originating near the center of the medium  and  
{\em fully penetrating} the plasma the enhanced escape time  
due to expansion $\tau=R/(1-v_T)$ compensates for the 
$1/(1+v_T\tau(\phi_0)/R)$ dilution factor. Therefore, in this isotropic case, 
the extra dilution due to transverse expansion 
 has in fact no effect of the total energy loss 
\beq
\Delta E^{(1)}_{1D}({\bf b}=0\; {\rm fm}) 
\approx \Delta E^{(1)}_{3D}({\bf b}=0\; {\rm fm})
\eeq{aprind}
modulo logarithmic  factors.  
We note that in performing the line integral Eq.~(\ref{deaz}), 
the  logarithm dependence on the cut-off, $\omega(\phi_0)$ was neglected.   
For our numerical estimates we  
approximate $\omega(\phi_0)$ by its average.
Our  first important conclusion is that the 
inclusive azimuthally averaged jet quenching pattern in central 
collisions is approximately independent of transverse expansion.

It is important to verify Eq.~(\ref{aprind}) for more realistic
density profiles since it implies that the overall suppression of the
high $p_T$ particle spectra is not affected by transverse flow.
We have checked numerically that this approximate 
line integral Eq.~(\ref{de11int}) independence of 
the transverse expansion velocity $v_T$ holds  for 
transverse density profiles with diffuse edges, namely a Gaussian profile 
\beqar  
\rho(\tau,\phi_0;x_0,y_0)&=&  
\frac{dN^g}{dy}\, \left(\frac{1}{\tau} \right)\, 
\frac{ 1 } {\pi (R^{eff}_x+v_x \tau) (R^{eff}_y+v_y \tau)}   \nonumber
  \\[1ex]  
&& \hspace{-1.3cm}
\exp \left[ -   \frac{(x_0+ \tau\cos \phi_0)^2}{(R^{eff}_x+v_x \tau )^2} - 
\frac{(y_0+ \tau \sin \phi_0)^2}{(R^{eff}_y+v_y \tau)^2} \right]   
\eeqar{rhogauss}  
and a curious  ($\phi_0$ asymmetric) exponential profile
\beqar  
\rho(\tau,\phi_0;x_0,y_0)&=&  
\frac{dN^g}{dy}\, \left(\frac{1}{\tau} \right)\, 
\frac{ 1 } {4 (R^{eff}_x+v_x \tau ) (R^{eff}_y+v_y \tau)}   \nonumber
  \\[1ex]  
&& \hspace{-1.3cm}
\exp \left[ -   \frac{|x_0+ \tau \cos \phi_0|}{(R^{eff}_x+v_x \tau )} - 
\frac{|y_0+\tau \sin \phi_0|}{(R^{eff}_y+v_y \tau)} \right]   \;.
\eeqar{rhoexpon}

One of the most interesting cases arises as a solution 
of Vlasov equation for a free-streaming massive partons. 
We consider the ideal Bjorken case  with perfect correlation 
between kinetic and space-time rapidities:
 $y=\eta$ with $y=\tanh^{-1}(p_z/p_0),\; \eta=\tanh^{-1}( z/t) $ and
transverse propagation at mid-rapidity ($\tau=t$).
Given initial transverse density $\rho_0(\vec{\bf x})$ 
and with azimuthally isotropic angular distribution
of plasma particles with a fixed magnitude of velocity, at time
$\tau$ one gets 
\beqar
 \rho(\vec{\bf x},\tau)&=&\frac{\tau_0}{\tau}\int \frac{d\Omega_{v_T}}
{2 \pi} \rho_0(\vec{\bf x}-\vec{v}_T \tau) \;\;.
\eeqar{isoexp}
For Gaussian initial density the evolution is actually given by 
\beqar  
\rho(r,\phi_0, \tau)&=&    
\frac{dN^g}{dy}\left(\frac{1}{\tau }\right) 
\frac{1}{\pi R^{eff}_x R^{eff}_y}   
  \int \frac{d \Omega_{v_T}}{2\pi}  
 \nonumber \\[1ex]  
&\;&\hspace{-2cm} \exp \left[-\frac{(r \cos\phi_0 - 
v_T \tau \cos\phi)^2}{(R_x^{eff})^2}   
-\frac{(r \sin\phi_0 - 
v_T \tau  \sin\phi)^2}{(R_y^{eff})^2}\right] \;. \nonumber \\[1ex]
%
\eeqar{vlasov}
This solution is rather different from Eq.~(\ref{rhogauss}) because the 
density expands as a thick shell
rather than homogeneously. The same observation applies for the sharp elliptic
profile Eq.~(\ref{rhoex}).
The general line integral that is needed assuming free Vlasov expansion
but averaging over initial coordinates and the angle of the internal particle
motion 
for fixed azimuthally isotropic $v_T$ is
\beqar
L(\hat{n})&=&\int d^2\vec{\bf x}_0 \; \tau_0 \rho_0(\vec{\bf x}_0,\tau_0) 
\int_{\tau_0}^\infty d\tau \; \tau  \rho (\vec{\bf x}_0+
\hat{n} \tau,\tau)
\nonumber \\[1ex]
&=& \tau_0 \int d^2 \vec{\bf x}_0 \; \tau_0 \rho_0(\vec{\bf x}_0,\tau_0) 
\int \frac{d\Omega_{v_T}}{2\pi} \nonumber \\[1ex] 
&& \qquad \int_0^\infty d\tau \;  
\rho_0(\vec{\bf x}_0+(\hat{n}-\vec{v}_T)\tau, \tau_0) \;\;. 
\eeqar{ave}
The explicit averaging over the jet production point $\vec{\bf  x}_0$
in this equation can only be performed numerically.  However for a jet
originating near the center of the medium  one can derive expression similar to
Eq.~(\ref{debj1})
\beqar
\Delta E^{(1)}_{\rm Vlas.\; 3+1D }(\phi_0)   
 &\approx& 
 \frac{9}{4} \, \frac{ C_R\alpha_s^3}{(R^{eff})^2}\frac{dN^g}{dy}  
\; \log \frac{E}{\omega(\phi_0)}  
\int_0^\infty d\tau  \nonumber \\[1ex]
&& \hspace{-1cm} \int \frac{d \phi}{2\pi}  \;
\exp\left( - \tau^2 \frac{1- 2 v_T \cos \phi + v_T^2}{(R^{eff})^2}    
    \right) \nonumber \\[1ex]
&& \hspace{-2cm}= \;\frac{9}{4} \, 
\frac{ C_R\alpha_s^3}{(R^{eff})^2}\frac{dN^g}{dy}  \;
\tau(\phi_0)_{\rm Vlas.}
\; \log \frac{E}{\omega(\phi_0)}  
\;\; . 
\eeqar{vlasexp}
The {\em equivalent escape factor} (enhanced propagation length
 modulated 
by the dilution of the medium) is given in this  Vlasov scenario
by  
$$\tau(\phi_0)_{\rm Vlas.} = 
\, \frac{R^{eff}}{\sqrt{\pi}\,(1+v_T)}\;
 K\left(\frac{4 v_T}{(1+v_T)^2} \right) \;\;,$$
where $K(x)$ is the complete elliptic integral of first kind. The
escape factor stays within $\pm10\%$ of $R^{eff}$ for flow velocities up to
$v_T=0.8$. As $v_T\rightarrow 1$, $\tau(\phi_0)_{\rm Vlas.}$ diverges
 only logarithmically.  
In case of non-central collisions, ${\bf b} \ne 0$,  and averaging over 
the jet production points, $\left\langle \Delta E^{(1)}  
\right\rangle_{\vec{\bf x}_0}$,  the validity of 
Eqs.~(\ref{aprind},\ref{vlasexp}) can be confirmed numerically.

In non-central collisions, the azimuthal asymmetry of the energy loss can be 
expanded in Fourier series and characterized as 
\beqar  
\Delta E^{(1)}_{3D}(\phi)&=&\Delta E(1+ 2 \delta_2(E)\cos 2\phi +\cdots)   
\; .   
\eeqar{debjv2}  
From Eq.~(\ref{deaz}), we compute numerically the azimuthally 
averaged mean $\Delta E$ 
and the energy loss asymmetry $\delta_2(E)$ averaging over 
the uniform elliptic 
density, Eq.~(\ref{ellip}), of the jet formation points.  
We consider only the specific case of a $E=10$ GeV gluon jet 
to illustrates the dependence of these quantities on the mean expansion  
velocity.  The spatial anisotropy is fixed by 
$R_x=2.5$~fm and $R_y=4.7$~fm that approximately corresponds 
 to the same second moment  
of the spatial anisotropy of the sharp intersecting cylinders considered 
in Ref.~\cite{gvw} for ${\bf b}=7$~fm. The  gluon plasma rapidity  
density that fixes  
the magnitude of the energy loss  
was assumed to be  $dN^g/dy\, ({\bf b}=7\;{\rm fm})=325$ 
with  $\alpha_s = 0.2$ for illustration. 

%
We calculate the energy loss and its asymmetry also by using a full
hydrodynamic calculation of Ref.~\cite{pasi}. In this case we use the
parametrization eBC of Ref~\cite{pasi} to initialize the system and treat
gluon number as conserved current to calculate the density evolution needed
in the line integral Eq.~(\ref{de11int}). We again average over the jet
formation points but in this case their density is not constant but given by
the number of binary collisions per unit area as in the Woods-Saxon geometry
of Ref~\cite{gvw}.
%

We plot in Figs.~1 and~2 the dependence of $\Delta E$ and 
$\delta_2(E=10\;{\rm GeV})$  on the mean transverse expansion 
velocities, fixing the velocity asymmetry from hydrodynamic results to be 
$v_x=v_T(1+\nu_a)$ and $v_y=v_T(1-\nu_a)$. Boosted thermal sources
simulations and hydrodynamic calculations indicate  
that $v_T \approx 0.5 - 0.6$~\cite{Adler:2001nb,Xu:2001zj,pasinew} 
and $\nu_a \approx 0.05 - 0.10$. 
The analytic computation and the hydro simulation in Fig.~1 were normalized 
to the same  mean energy loss for the $v_T=0$ point.  
We see that $\Delta E$ is weakly dependent on transverse flow for the
schematic but analytically tractable uniform expanding elliptic density and
we have obtained similar results for a wide variety of transverse profiles
Eqs.~(\ref{rhogauss},\ref{rhoexpon},\ref{vlasov}).   
Full  hydrodynamic  calculation of the $\phi$ and $\vec{\bf x}_0$
averaged line integral Eq.~(\ref{de11int}) is also consistent with  
approximately constant mean energy loss within $\pm5\%$ for both
central and semi-central collisions.

However, the  azimuthal asymmetry of the energy loss is strongly 
reduced for realistic hydrodynamic flow velocities. This implies a 
much smaller $v_2$ at high $p_T$ 
than obtained in Ref.~\cite{gvw} where transverse 
expansion was not considered. 
Hydrodynamic calculations with transverse expansion predict similar 
reduction  of the  averaged line integral Eq.~(\ref{de11int}). 
We note that factor  of 1.5 - 2 difference in $\delta_2(E)$ 
without transverse flow in Fig.~2 reflects 
the difference between the sharp cylinder geometry and the Wood-Saxon 
geometry as shown in Ref.~\cite{gvw}.

%
%
The error bars in the hydro calculation are our estimate for uncertainties in
terminating the line integral Eq.~(\ref{de11int}) at freeze-out and the
behavior of density during the mixed phase and hadronization.
The correspondence of the expansion velocity $v_T$ in hydro and 
schematic model is also ambiguous since in hydro $v_T$ is the average 
velocity on the freeze-out surface, not the velocity at the edge of 
the system as in Eq.~(\ref{ellip}).
%

\begin{center} 
\vspace*{8.5cm} 
\includegraphics{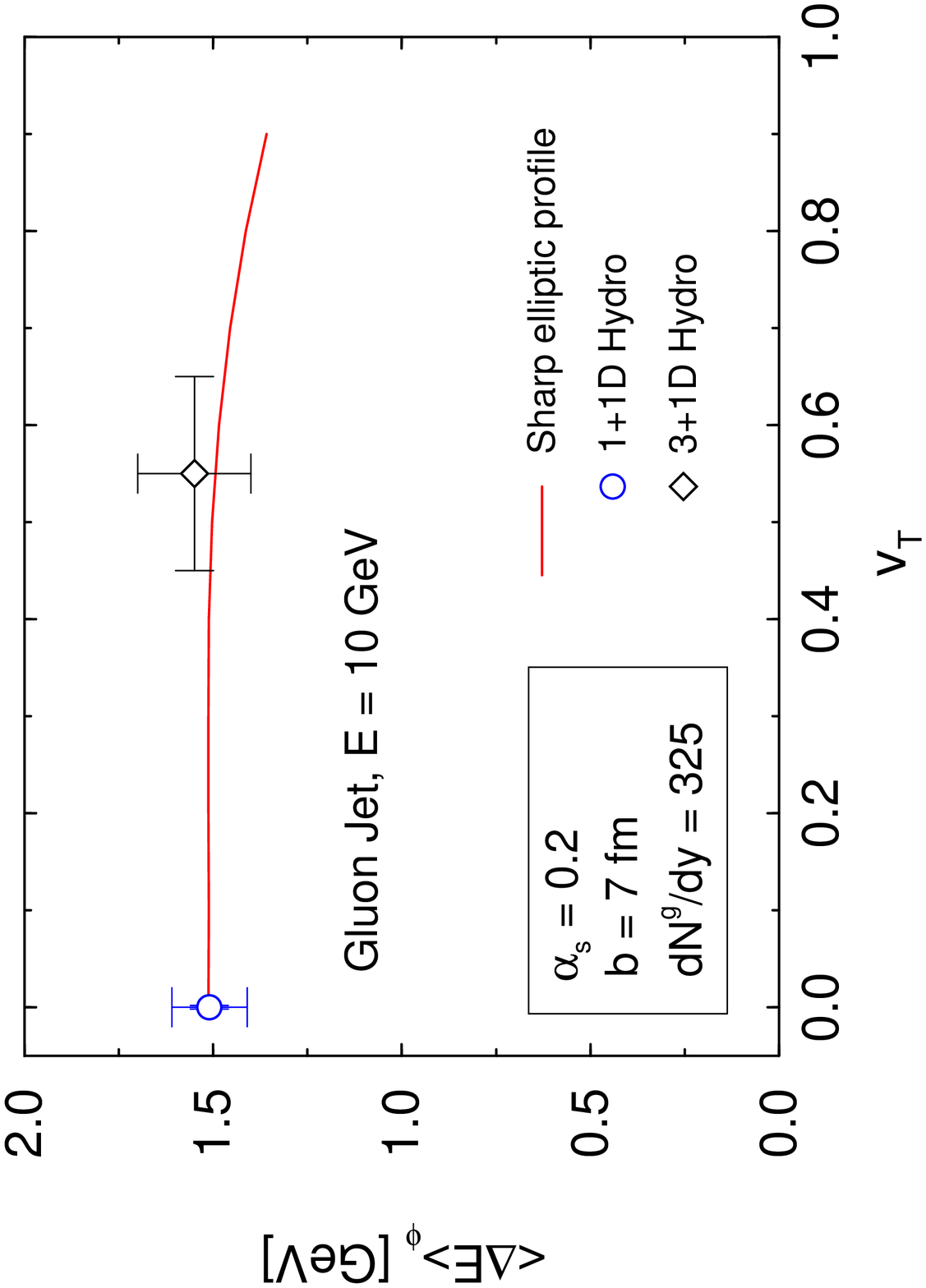} 
\vspace{-2.9cm} 
\end{center} 
\begin{center} 
\begin{minipage}[t]{8.5cm} 
         { FIG. 1.} {\small   
The azimuthally averaged energy loss for a 10~GeV gluon jet  
propagating through a 3+1D elliptic expanding plasma is plotted 
as a function of the mean 
transverse flow velocity, $v_T$. The initial profile is a homogeneous ellipse 
approximating $Au+Au$ at ${\bf b}=7$~fm impact parameter. Transverse velocity 
asymmetry is fixed to be $\nu_a=0.1$. 1+1D and 3+1D hydro calculations 
are shown.
} 
\end{minipage} 
\end{center}

\begin{center} 
\vspace*{8.5cm} 
\includegraphics{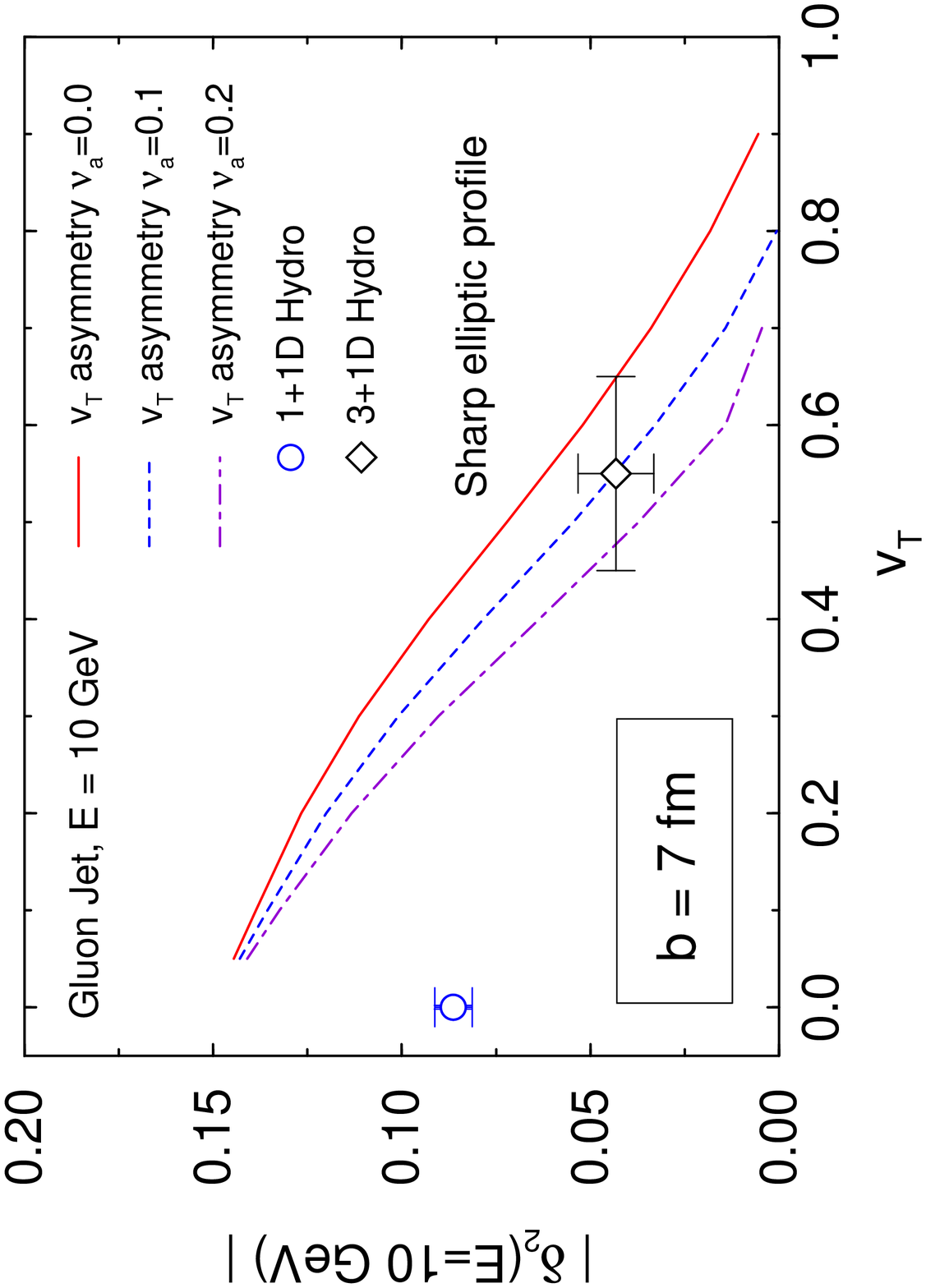} 
\vspace{-2.6cm} 
\end{center} 
\begin{center} 
\begin{minipage}[t]{8.5cm} 
         { FIG. 2.} {\small   
The second harmonic of the jet energy loss 
for a 10~GeV gluon 
propagating through a 3+1D elliptic expanding plasma as a function of the mean 
transverse flow velocity, $v_T$, is shown. 
The initial profile is chosen as in Fig.~1. 
We study 3 different transverse velocity  
asymmetries, i.e. $\nu_a=0.0,\; 0.1,\; 0.2$. Hydrodynamic 1+1D and 3+1D
calculations using diffuse Wood-Saxon geometry are also included. 
} 
\end{minipage} 
\end{center}

\section{High $p_T$ Azimuthal Asymmetry Scenarios} 
 
The results of the previous section 
suggest that if  strong transverse expansion of the plasma  
can be confirmed (and other scenarios 
including soft string fragmentation, 
baryon junctions~\cite{gv,gvproc,nestor} and classical 
Yang-Mills~\cite{raju} evolution rejected) 
then the high transverse momentum asymmetry 
predicted by energy loss calculations may be  reduced (by a factor 
of 2 to 4). This would leave open the understanding 
of the present STAR data in terms of the eikonal picture~\cite{gvw}. 
Interesting new possibilities  arise 
to explain  the moderate to high 
$p_T$ dependence of the transverse azimuthal asymmetry 
$v_2(p_T>2\;{\rm GeV/c})$. 
We suggest that the key to converging to the correct dynamical picture 
is the flavor dependence of both the asymmetry and the $p_T$ differential 
particle  multiplicities. 
We discuss three possible scenarios for high $p_T$  hadron production 
with expected $v_2(p_T)$  behavior shown in Fig.~3.   
\begin{center} 
\vspace*{8.5cm} 
\includegraphics{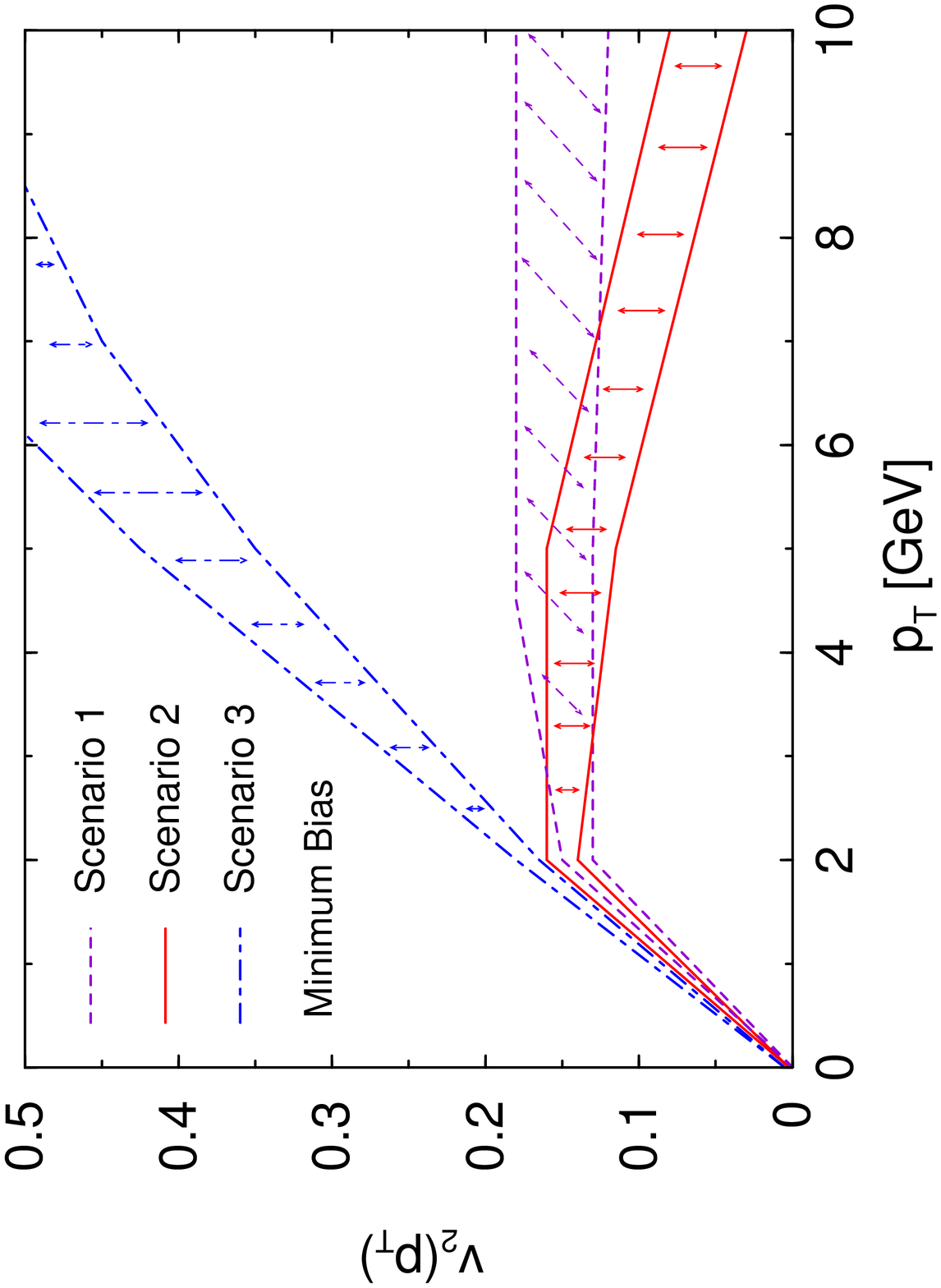} 
\vspace{-3.0cm} 
\end{center} 
\begin{center} 
\begin{minipage}[t]{8.5cm} 
         { FIG. 3.} {\small   
Three possible scenarios of the $v_2(p_T)$ behavior  are shown.
Error bands are suggestive of the theoretical uncertainties  
within each framework (e.g. the density of the medium or the transverse 
expansion velocity $v_T$).
} 
\end{minipage} 
\end{center}

\begin{enumerate} 
\item    
If experiments confirm an approximate 
 saturation of  $v_2 \sim 0.10-0.15$ at  high $p_T$  
for {\em both} pions and protons  {\em together} with a baryon to meson  
ratio dropping back below unity this may suggest that  
the transverse flow  might be
significantly smaller than the values used
in the thermal model description of relativistic heavy ion collisions.
The assumption of very large transverse expansion velocities 
may be a caveat since it is not supported by pion interferometry
as pointed out in Ref.~\cite{STARint}.  
In this case the energy loss  asymmetry  plays significant 
role~\cite{gvw}, producing comparable
asymmetries for $p$ and $\pi$. 
Alternatively, appropriate description of particle production 
could  be  provided  by dissipative relativistic hydrodynamics~\cite{muro} 
or inelastic parton cascades~\cite{denes} with  large partonic 
cross sections  $\sigma \sim  10 - 40$~mb. 
 
\item 
Another possibility arises if the pion multiplicity at moderate  high 
transverse momenta is dominated by the quenched pQCD spectra 
but the baryon production at moderate transverse momenta $2<p_T<5$~GeV/c
 is controlled by a non-perturbative  mechanism such as baryon junctions 
(which exhibit azimuthal asymmetry and flow) 
or hydro (with baryochemical potential which 
can be inferred from the baryon transport picture).  
A distinctive characteristic of this picture is that 
one expects qualitatively  different $v_2(p_T)$ 
behavior for $p$ and $\pi$ and a baryon/meson anomaly 
limited to the intermediate transverse momentum 
window~\cite{gv,gvproc,gvPetproc}. 
Above $p_T=2$~GeV/c the proton and 
anti-proton $v_2$ may keep growing due to flow effects, whereas the pion  
$v_2$, being driven by asymmetrically quenched pQCD, may start decreasing 
because strong transverse expansion (if confirmed)  
reduces the energy 
loss azimuthal anisotropy. The comparable 
contribution of baryons and mesons in this 
region~\cite{bill,gv,gvproc,gvPetproc,julia}   
leads to cancelation of those effects in the inclusive charged particle 
elliptic flow, thus producing a plateau in the  $2<p_T<5$~GeV/c window.  
At high transverse momenta of $p_T \sim 6-10$~GeV/c the pQCD  
pions dominate the charged particle multiplicity, 
thus leading to a gradual decrease of $v_2$. The 
rate at which the high $p_T$ azimuthal asymmetry vanishes is determined 
by the  mean transverse expansion velocity $v_T$, the velocity asymmetry 
(see Fig.~2), and the pQCD fractional contribution to the baryon  
differential $p_T$ multiplicities. 
The {\em schematic} $v_2$ behavior in this scenario is illustrated 
in Fig.~4. It  predicts a detectable difference in the moderate to high 
transverse momentum behavior of the elliptic flow for pions and protons.

\item 
Last, but not least, there are  predictions based on 
boosted thermal sources and 
ideal hydrodynamics. The predicted $v_2(p_T)$ grows 
continuously as a function of the transverse momentum and 
saturates at a maximum value. 
This was parametrized from hydro simulations~\cite{kolb} in
Ref.~\cite{gvw} by
$$v_2(p_T)= \tanh[p_T/(10 \pm 2 \; {\rm GeV/c})] \;\;.$$
Similarly the baryon/meson ratio (assuming equal 
freeze-out temperatures) 
is a monotone function of $p_T$ and saturates at $(p/\pi)_{\rm max}=2$ from 
simple spin counting. 
 
\end{enumerate} 

Precision data on identified hadron spectra  
at high $p_T$ at  RHIC  is needed to 
shed light on which if any of these dynamical scenarios 
predicts the correct flavor dependence of the azimuthal asymmetry 
and the proton to pion ratio.

\begin{center} 
\vspace*{8.5cm} 
\includegraphics{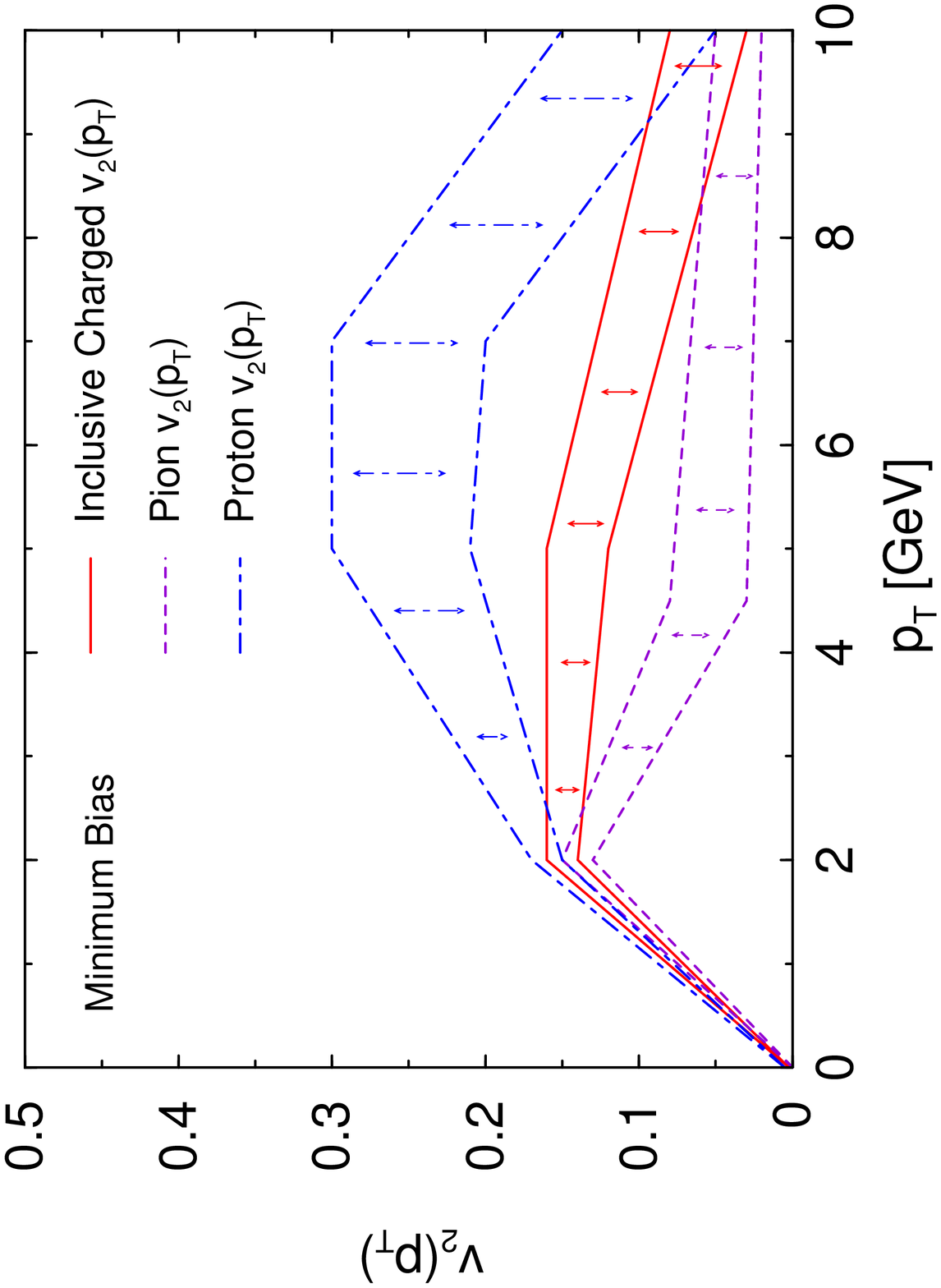} 
\vspace{-3.0cm} 
\end{center} 
\begin{center} 
\begin{minipage}[t]{8.5cm} 
         { FIG. 4.} {\small   
Qualitative illustration of the $v_2(p_T)$ behavior for pions, protons and 
inclusive charged particles within the framework of the second 
scenario is given. 
A characteristic difference in the elliptic flow of $p$ and $\pi$ at moderate 
high $p_T$ is illustrated. Error bands hint at uncertainties related to the 
time for developing a rarefaction wave, flow velocities, pQCD vs. hydro vs.
string and junction components  and the baryon/meson flavor composition of
charged hadrons.
} 
\end{minipage} 
\end{center}

\section{Conclusions} 
We have analyzed the effect of possible strong transverse flow 
$v_T\sim 0.6$~\cite{Xu:2001zj,gv,gvproc}  
 on  the transverse azimuthal asymmetry generated by jet 
energy loss. We have shown  that while the overall magnitude of the 
quenching is only weakly dependent 
on $v_T$, the azimuthal asymmetry 
$v_2$ of the quenched pQCD component is significantly reduced.  
In order to account for the observed saturation of $v_2$ of {\em charged} 
hadrons, scenario 2 based on~\cite{gvw,gv}  
is forced into  predicting an anomalous asymmetry of the 
baryon component alone. This scenario interprets  
the anomalous baryon to pion ratio in terms of baryon junction  
dynamics~\cite{gv,gvproc,gvPetproc}. 
It differs from scenarios 1 and 3 in predicting that the pion asymmetry  
is expected to vanish more quickly with increasing $p_T$, 
 while the baryon asymmetry is expected to be near maximal up 
to $\sim 4-7$ GeV/c. 
Hydrodynamic and dissipative transport models expect 
a much less dramatic  dependence on $p_T$.

\acknowledgements 
We thank Adrian Dumitru and Denes Molnar 
for helpful discussion.
This work was supported by the Director, Office of Science,  
Office of High Energy and Nuclear Physics, 
Division of Nuclear Physics  of the U.S. Department of Energy 
under Contract Nos. DE-FG-02-93ER-40764 and DE-AC03-76SF00098.

 
\vfill\eject 
\end{multicols} 

\begin{references} 
 
 
\bibitem{Snellings} 
R.J.~Snellings  [STAR Collaboration], 
nucl-ex/0104006. 
 
 
\bibitem{starv2} 
K.H.~Ackermann {\it et al.}  [STAR Collaboration], 
Phys. Rev. Lett. {\bf 86}, 402 (2001).  
[nucl-ex/0009011]. 
 
\bibitem{Adler:2001nb} 
C.~Adler {\it et al.}  [STAR Collaboration], 
nucl-ex/0107003. 
 
 
\bibitem{reanv2} 
A.H.~Tang [STAR Collaboration], 
hep-ex/0108029.  
 
 
 
\bibitem{gvw} 
M.~Gyulassy, I.~Vitev, X.-N.~Wang, Phys. Rev. Lett. {\bf 86}, 2537 (2001).  
 
 
\bibitem{wangv2} 
X.-N.~Wang, Phys. Rev. {\bf C63}, 054902 (2001).   
 
 
\bibitem{glvprl} M.~Gyulassy, P.~L\'evai, I.~Vitev, 
Phys. Rev. Lett. {\bf 85}, 5535 (2000).  
 
\bibitem{glv2} M.~Gyulassy, P.~L\'evai and I.~Vitev, 
Nucl. Phys.  {\bf B594}, 371  (2001).  
 
\bibitem{glv1b} M.~Gyulassy, P.~L\'evai, I.~Vitev,
Nucl. Phys. {\bf B571}, 197 (2000). 

\bibitem{glv1a} M.~Gyulassy, P.~L\'evai, I.~Vitev,
Nucl. Phys. {\bf A661}, 637c (1999).
 



 
\bibitem{pi0} 
G.~David, [PHENIX Collaboration], nucl-ex/0105014.  
 
\bibitem{bill} 
W.A.~Zajc, [PHENIX Collaboration], nucl-ex/0106001. 

\bibitem{phenqnch} 
K.~Adcox {\em et al.}, [PHENIX Collaboration], 
nucl-ex/0109003, submitted to PRL. 
 
\bibitem{pasi} 
P.F.~Kolb, U.~Heinz, P.~Huovinen, K.J.~Eskola, K.~Tuominen, 
hep-ph/0103234.  

\bibitem{Hirano:2001yi}
T.~Hirano, K.~Morita, S.~Muroya and C.~Nonaka,
nucl-th/0110009.

 
\bibitem{Adler:2001yq} 
C.~Adler {\it et al.}  [STAR Collaboration], 
nucl-ex/0106004. 
 
\bibitem{Xu:2001zj} 
N.~Xu and M.~Kaneta, 
nucl-ex/0104021. 

\bibitem{Broniowski:2001we}
W.~Broniowski and W.~Florkowski,
nucl-th/0106050.


\bibitem{gv} 
I.~Vitev, M.~Gyulassy, nucl-th/0104066.  
 
\bibitem{gvproc} 
I.~Vitev, M.~Gyulassy,  hep-ph/0108045.  


\bibitem{gvPetproc} 
I.~Vitev, M.~Gyulassy, P.~L\'evai, To appear in EPS HEP2001 proceedings.  


\bibitem{pasiold} 
J.~Sollfrank, P.~Huovinen, M.~Kataja, P.V.~Ruuskanen, M.~Prakash, 
R.~Venugopalan, Phys. Rev. {\bf C55}, 392 (1997). 

\bibitem{pasinew} 
P.~Kolb, nucl-th/0104089. 

\bibitem{zakh} 
B.~Zakharov, 
JETP Lett. {\bf 73}, 49 (2001). 
 
\bibitem{levai} 
P. Levai, G. Papp, G. Fai, M. Gyulassy, nucl-th/0012017. 

\bibitem{WW} E.~Wang, X.-N. Wang, nucl-th/0106043. 
 
\bibitem{nestor} 
N.~Armesto, C.~Pajares, D.~Sousa, 
hep-ph/0104269.  
  
\bibitem{raju} 
A.~Krasnitz, Y.~Nara, R.~Venugopalan, 
hep-ph/0108092.  

\bibitem{STARint} 
C.~Adler {\em et al.},
Phys. Rev. Lett. {\bf87}, 082301 (2001). 
 
\bibitem{muro} 
A.~Muronga,  
nucl-th/0104064.  
 
\bibitem{denes} 
D.~Molnar, M.~Gyulassy,  
nucl-th/0104073.  
  
\bibitem{julia} 
J.~Velkovska [PHENIX Collaboration], nucl-ex/0105012.  
 
\bibitem{kolb} 
P.F.~Kolb, P.~Huovinen, U.~Heinz, H.~Heisenberg,  
Phys. Lett. {\bf B500}, 232 (2001).
  
\end{references}
\end{document}